%% file: readability.tex
\documentclass[runningheads]{llncs}
\usepackage{times}
\usepackage{setspace}
\usepackage[breaklinks=true,hidelinks=true]{hyperref}
\usepackage{color}
\usepackage{comment}
\usepackage{url}

\usepackage{booktabs}
\usepackage{amsfonts}
\usepackage{amsmath}
\usepackage{amssymb}
\usepackage{makecell}
\usepackage{subcaption}
\usepackage{multirow}
\usepackage{pbox}

\usepackage[flushleft]{threeparttable}

\usepackage{graphicx,xspace}
\usepackage{epstopdf}
\usepackage{stfloats}
\usepackage[font=small]{caption}
\usepackage{enumitem}
\usepackage{wrapfig}
\usepackage{flushend}
\def\st#1{~}

\input{math_commands}

\begin{document}
%
\title{ReadNet: A Hierarchical Transformer Framework for Web Article Readability Analysis}
\titlerunning{ReadNet: A Hierarchical Transformer Framework for Readability Analysis}
\author{
Changping Meng$^1$\thanks{This work was done during the summer internships of CM and MC at Google, Mountain View. We thank the anonymous reviewers for their insightful comments.}, Muhao Chen$^2$, Jie Mao$^3$,Jennifer Neville $^1$
}
\institute{
Department of Computer Science, Purdue University, West Lafayette$^1$\\
Department of Computer Science, University of California, Los Angeles$^2$\\
Google Inc., Mountain View$^3$\\
\{meng40, neville\}@purdue.edu; muhaochen@ucla.edu; mjmjmtl@google.com
}

\maketitle
\input{abstract}
\input{introduction}
\input{related}
\input{model}
\input{experiment}
\input{conclusion}

\newpage
\bibliographystyle{splncs04}
\input{reference}

\end{document}

%% file: math_commands.tex
\usepackage{amsmath,amsfonts,bm}

\makeatletter
\let\overlinewithoriginalheight\overline
\newcommand*\overlinewithlessheight[1]{{\mathpalette\overline@aux{#1}}}
\newcommand*\overline@aux[2]{
  \begingroup
    \count0=\fam 
    \setbox0=\hbox{$\m@th #1\fam=\count0 #2$}
    \@tempdima=.4\ht0
    \setbox0=\hbox{$\m@th #1\fam=\count0\overlinewithoriginalheight{#2}$}%
    \advance\@tempdima by .6\ht0
    \ht0=\@tempdima 
    \usebox0
  \endgroup%
}
\let\overline\overlinewithlessheight

\let\underlinewithoriginaldepth\underline
\newcommand*\underlinewithlessdepth[1]{{\mathpalette\underline@aux{#1}}}
\newcommand*\underline@aux[2]{%
  \begingroup
    \count0=\fam
    \setbox0=\hbox{$\m@th #1\fam=\count0 #2$}%
    \@tempdima=.4\dp0%
    \setbox0=\hbox{$\m@th #1\fam=\count0\underlinewithoriginaldepth{#2}$}%
    \advance\@tempdima by .6\dp0%
    \dp0=\@tempdima
    \usebox0%
  \endgroup%
}
\let\underline\underlinewithlessdepth
\makeatother

\newcommand{\harrow}[1]{\mathstrut\mkern2.5mu#1\mkern-11mu\raise1.6ex\hbox{$\scriptscriptstyle\rightharpoonup$}}

\def\1{\bm{1}}

\def\vh{{\bm{h}}}

\def\vr{{\bm{r}}}

\def\vu{{\bm{u}}}
\def\vv{{\bm{v}}}
\def\vw{{\bm{w}}}

\def\vy{{\bm{y}}}

\def\mA{{\bm{A}}}
\def\mB{{\bm{B}}}
\def\mC{{\bm{C}}}

\def\mH{{\bm{H}}}

\def\mK{{\bm{K}}}

\def\mP{{\bm{P}}}
\def\mQ{{\bm{Q}}}

\def\mU{{\bm{U}}}
\def\mV{{\bm{V}}}
\def\mW{{\bm{W}}}
\def\mX{{\bm{X}}}

\DeclareMathAlphabet{\mathsfit}{\encodingdefault}{\sfdefault}{m}{sl}
\SetMathAlphabet{\mathsfit}{bold}{\encodingdefault}{\sfdefault}{bx}{n}



%% file: abstract.tex
\begin{abstract}
Analyzing the {\em readability} of articles has been
an important sociolinguistic task.
Addressing this task is necessary to the automatic recommendation of appropriate articles to readers with different comprehension abilities, and it further benefits education systems, web information systems, and digital libraries.
Current methods for assessing readability employ empirical measures or statistical learning techniques that are limited by their ability to characterize complex patterns such as article structures and semantic meanings of sentences.
In this paper, we propose a new and comprehensive framework which uses a hierarchical self-attention model to analyze document readability.
In this model, measurements of sentence-level difficulty are captured along with the semantic meanings of each sentence.
Additionally, the sentence-level features are incorporated to characterize the overall readability of an article with consideration of article structures.
We evaluate our proposed approach on three widely-used benchmark datasets against several strong baseline approaches. Experimental results show that our proposed method achieves the state-of-the-art performance on estimating the readability for various web articles and literature.
\end{abstract}

%% file: introduction.tex
\section{Introduction}\label{intro}
Readability is an important linguistic measurement that indicates how easily readers can comprehend a particular document.
Due to the explosion of web and digital information, there are often hundreds of articles describing the same topic, but vary in levels of readability.
This can make it challenging for users to find the articles online that better suit their comprehension abilities.
Therefore, an automated approach to assessing readability is a critical component for the development of 
recommendation strategies for web information systems, including digital libraries and web encyclopedias.

{\em Text readability} is defined as the overall effect of language usage and composition on readers' ability to easily and quickly comprehend the document~\cite{readability}.
In this work, we focus on evaluating document difficulty based on the composition of words and sentences.
Consider the following two descriptions of the concept \emph{rainbow} as an example.


\begin{enumerate}
\item \textbf{A more rigid scientific definition from \emph{English Wikipedia}}: A rainbow is a meteorological phenomenon that is caused by reflection, refraction and dispersion of light in water droplets resulting in a spectrum of light appearing in the sky. 
\item \textbf{A more generic description from the \emph{Simple English Wikipedia}}: A rainbow is an arc of color in the sky that can be seen when the sun shines through falling rain. The pattern of colors starts with red on the outside and changes through orange, yellow, green, blue, to violet on the inside.
\end{enumerate}
Clearly, the first description provides more rigidly expressed contents, but is more sophisticated due to complicated sentence structures and the use of professional words. In contrast, the second description is simpler, with respect to both grammatical and document structures.  From the reader's perspective, the first definition is more appropriate for technically sophisticated audiences, while the second one is suitable for general audiences, such as parents who want to explain rainbows to their young children.

The goal of \emph{Readability Analysis} is to provide a rating regarding the difficulty of an article for average readers.
As the above example illustrates that, many approaches for automatically judging the difficulty of the articles are rooted in two factors: the difficulty of the words or phrases, and the complexity of syntax~\cite{CollinsSurvey}. 
To characterize these factors, existing works~\cite{readability2,readability3} mainly rely on some explicit features such as \emph{Average Syllables Per Word}, \emph{Average Words Per Sentence}, etc. For example, the Flesch-Kincaid index is a representative empirical measure defined as a linear combination of these factors~\cite{chall1995readability}. Some later approaches mainly focus on proposing new features with the latest Coh\-Metrix 3.0~\cite{mcnamara2014automated} providing 108 features, and they combine and use the features using either linear functions or statistical models such as Support Vector Machines or multilayer perceptron~\cite{SVM,collins2005predicting,pitler2008revisiting,pilan2016predicting,CambridgeData}. While these approaches have shown some merits, they also lead to several drawbacks. Specifically (1) they do not consider sequential and structural information, and  (2) they do not capture sentences-level or document-level semantics that are latent but essential to the task~\cite{CollinsSurvey}.


To address these issues, we propose ReadNet, a comprehensive readability classification framework that uses a hierarchical transformer network. The self-attention portion of the transformer encoder is better able to model long-range and global dependencies among words. The hierarchical structure can capture how words form sentences, and how sentences form documents, meanwhile reduce the model complexity exponentially.
Moreover, explicit features indicating the readability of different granularities of text can be leveraged and aggregated from multiple levels of the model.
We compare our proposed model to a number of widely-adopted document encoding techniques,
as well as traditional readability analysis approaches based on explicit features.
Experimental results on three benchmark datasets show that our work properly identifies the document representation techniques,
and achieves the state-of-the-art performance by significantly outperform previous approaches. 

%% file: related.tex
\section{Related Work}\label{sec:related}
Existing computational methods for readability analysis~\cite{readability2,readability3,readability4,CollinsSurvey,pilan2016predicting} mainly use empirical measures on the symbolic aspects of the text, while ignoring the sequence of words and the structure of the article. The Flesch-Kincaid index~\cite{kincaid1975derivation} and related variations use a linear combination of explicit features.

Although models based on these traditional features are helpful to the quantification of readability for small and domain-specific groups of articles, they are far from generally applicable for a larger body of web articles~\cite{Si2001ASM,CollinsThompson2004ALM,Feng2009}. Because those features or formulas generated from a small number of training text specifically selected by domain experts, they are far from generally representing the readability of large collections of corpora. Recent machine learning methods on readability evaluation are generally in the primitive stage. \cite{franccois2009combining} proposes to combine language models and logistic regression. The existing way to integrate features is through a statistical learning method such as SVM~\cite{ReadingMeasures,SVM,collins2005predicting,pitler2008revisiting,pilan2016predicting,CambridgeData}. 
These approaches ignore the sequential or structural information on how sentences construct articles. Efforts have also been made to select optimal features from current hundreds of features~\cite{de2016all}. Some computational linguistic methods have been developed to extract higher-level language features. The widely-adopted Coh-Metrix~\cite{graesser2004coh,mcnamara2010coh} provides multiple features based on cohesion such as referential cohesion and deep cohesion.

Plenty of works have been conducted on utilizing neural models for sentimental or topical document classification or ranking, 
while few have paid attention to the readability analysis task.
The convolutional neural network (CNN)~\cite{CNN} is often adopted in sentence-level classification which leverages local semantic features of sentence composition that are provided by word representation approaches. In another line of approaches, a recursive neural network~\cite{socher2013recursive} is adopted, which focuses on modeling the sequence of words or sentences. 
Hierarchical structures of such encoding techniques are proposed to capture structural information of articles, and have been widely used in tasks of document classification~\cite{tang2015document,lin2015hierarchical,chen2019subarticle}, and sequence generation~\cite{li2015hierarchical} and sub-article matching \cite{chen2018neural}. Hierarchical attention network~\cite{RNN} is the current state-of-the-art method for document classification, which employs attention mechanisms on both word and sentence levels to capture the uneven contribution of different words and sentences to the overall meaning of the document.  The Transformer model~\cite{vaswani2017attention} uses multi-head self-attention to perform sequence-to-sequence translation. Self-attention is also adopted in text summarization, entailment and representation~\cite{parikh2016decomposable,li2018hierarchical}. 
Unlike topic and sentiment-related document classification tasks that focus on leveraging portions of lexemes that are significant to the overall meanings and sentiment of the document, readability analysis requires the aggregation of difficulty through all sentence components. Besides, precisely capturing the readability of documents requires the model to incorporate comprehensive readability-aware features, including difficulty, sequence and structure information, to the corresponding learning framework.

%% file: model.tex
\section{Preliminary}\label{sec:setting}

  In this section, we present the problem definition, as well as some representative explicit features that are empirically adopted for the readability analysis task.

  \subsection{Problem Definition}

  The readability analysis problem is defined as an ordinal regression problem for articles. Given an article with up to $n$ sentences and each sentence with up to $m$ words, an article can be represented as a matrix $\mA$ whose $i$-th row $\mA_{i,:}$ corresponds to the $i$-th sentences, and $A_{i,j}$ denotes the $j$-th word of the $i$-th sentence.
  Given an article $\mA$, a label will be provided to indicate the readability of this article.

We consider the examples introduced in Section~\ref{intro}, where two articles describe the same term ``\textit{rainbow}". The first rigorous scientific article can be classified as ``difficult", and the second general description article can be classified as ``easy". 

Instead of classifying articles into binary labels like ``easy" or ``difficult", more fine-grained labels can help people better understand the levels of readability. For instance, we can map the articles in standardization systems of English tests such as 5-level Cambridge English Exam (CEE),
where articles from professional level English exam (CPE) are regarded than those from introductory English exam (KET).

    %



  \subsection{Explicit Features} \label{indicator}
  Previous works~\cite{kincaid1975derivation,gibson1998linguistic,heilman2007combining,heilman2008analysis,malvern2012measures,CollinsSurvey,graesser2004coh} have proposed empirical features to evaluate readability. Correspondingly, we divide these features into sentence-level features and document-level features. Sentence-level features seek to evaluate the difficulty of sentences. 
  For instance, the sentence-level feature ``number of words" for sentences can be averaged into ``number of words per sentence" to evaluate the difficulty of documents. Document-level features include the traditional readability indices and cohesion's proposed by Coh-Metrix\cite{graesser2004coh}. These features are listed in Table~\ref{tb:features}.

  \begin{table*}[!h]
  	\centering
  	\scriptsize
  	\begin{tabular}{p{0.25\columnwidth}|p{0.75\columnwidth}}
  		{Name} & Description \\
  		\hline
  		\multicolumn{2}{l}{Sentence-level features}\\
  		\hline
  		\#characters\_per\_word &  The average number of characters per word, which provides a character-level measure for the difficulty of words. \\
  		\#syllabi\_per\_word & The average number of syllabi per word, which measures the difficulty of words from the syllabus level. \\
  		\#words& The number of words that measures the verbosity of the sentence. \\
  		\#long\_words &  The number of words longer than 6 characters in a sentence. \\
  	\#difficult\_words & The number of difficult word in a sentence. Difficult word is a word not listed in the 3000 words for fourth-grade American students.\\
  		\#pronoun & The number of pronoun in a sentence. \\
  		\hline
  		\multicolumn{2}{l}{Document-level features}  \\
  		\hline
  		Flesch Reading Ease~\cite{kincaid1975derivation} & The United States Military Standard of readability scoring for technical manuals, which is calculated as $206.835 - 1.015 \times \frac{\mbox{\#words}}{\mbox{\#sentences }} - 84.6 \times \frac{\mbox{\#syllables}}{\mbox{\#words}}$. \\
  		Flesch-–Kincaid grade level~\cite{kincaid1975derivation} & An empirical readability metric which maps to a U.S. school grade level, calculated as $0.39 \times \frac{\mbox{\#words }}{\mbox{\#sentences  }} + 11.8 \times \frac{\mbox{\#syllables }}{\mbox{\#words }} - 15.59$.\\
  		Automated Readability Index~\cite{senter1967automated} & A metric that also produces an approximate representation of the US grade level needed to comprehend the text, calculated as $ 4.71 \times \frac{\mbox{\#characters}}{\mbox{\#words}} + 0.5  \times \frac{\mbox{\#words}}{\mbox{\#sentences }}$ - 21.43. Instead of considering syllables, this metric more generally characterizes on the character level.\\
  		Coleman-Liau Index~\cite{coleman1975computer} & An index used to gauge the understandability of a text from the character-level: $0.0588 \times \frac{\mbox{\#letters}}{\mbox{\#words} \times 100} + 0.296 \times \frac{\mbox{\#sentences}}{\mbox{\#words}} \times 100 $.\\
  		Gunning Fog Index~\cite{gunning1969fog}& $0.4 \times (\frac{\mbox{\#words}}{\mbox{\#sentences }} + 100 \times \frac{\mbox{\#complex\_words}}{\mbox{\#words}})$; It estimates the years of formal education a person needs to understand the text on the first reading.\\
  		LIX~\cite{brown2005student} & A measure indicating the difficulty of reading a text based on the proportions of long words and verbosity of sentences: $\frac{\mbox{word longer than 6 letters \#}}{\mbox{\#words}} +   \frac{\mbox{\#words}}{\mbox{\#sentences }} $.\\
		RIX~\cite{anderson1983lix} & A metric based on the proportion of long words in text, $\frac{\mbox{\# long            words}}{\mbox{\#sentences }}$.\\
  		SMOG Index~\cite{mc1969smog} & A measure of readability that seeks to estimate the years of education needed to understand a piece of writing: $1.0430 \times \sqrt{\mbox{\# of polysyllables}\times{30 \over \mbox{\#sentences }} } + 3.1291$.\\
  		Dale Chall Index~\cite{fry1968readability} & $0.1579 \times \frac{\mbox{\#difficult\_words}}{\mbox{\#words}}\times 100  + 0.0496 \times \frac{\mbox{\#words}}{\mbox{\#sentences }}$. Difficult word is a word not listed in the 3000 words for fourth-grade American students\\
  		Incidence of connectives~\cite{louwerse2001analytic} & 5 numerical features indicate additive, logic, temporal, causal and
negative connectives.\\
  		Logic operator connectivity~\cite{coxhead2000new} & Logical connectives between logical particles such as ``and'', ``if'' proposed by Coh-Metrix. \\
  		Lexical diversity & The character-level density of the lexicon: $\frac{\mbox{\#unique\_words}}{\mbox{\#words}}$. \\
  		Content diversity &  $\frac{\mbox{\#content\_words }}{\mbox{\#words}}$. It measures the diversity of content. Content words are adjectives, nouns, verbs and adverbs.\\
  		Incidence of part-of-speech elements & Incidence of word categories (adjectives, nouns, verbs, adverbs, pronouns) per 1000 words
in the text \\
  	\end{tabular}
  	\caption{Explicit Features}\label{tb:features}
  	\vspace{-1cm}
  \end{table*}
Current approaches~\cite{SVM,collins2005predicting,pitler2008revisiting} average the sentence-level features of each sentence to construct document level features.
Furthermore, these features are concatenated with document-level features, and use an SVM to learn on these features. The limitation lies in failing to capture the structure information of sentences and documents. For instance, in order to get the sentence level features for the document, it averages all these features of each sentence. It ignores how these sentences construct an article and which parts of the document more significantly decides the readability of the document. While cohesion features provided by Coh-Metrix tries to captures relationships between sentences, these features mainly depend on the repeat of words across multiple sentences. They did not directly model how these sentences construct a document in perspectives of structure and sequence.

Briefly speaking, existing works are mainly contributing more features as shown in Table~\ref{tb:features}. But the current models used to aggregate these features are based on SVM and linear models. In this work, we target to propose a more advanced model to better combine these features with document information.

  \section{Hierarchical Transformer for Readability Analysis}
\begin{figure}[!ht]
\centering
    \vspace{-0.5cm}
	\includegraphics[width=0.90\columnwidth]{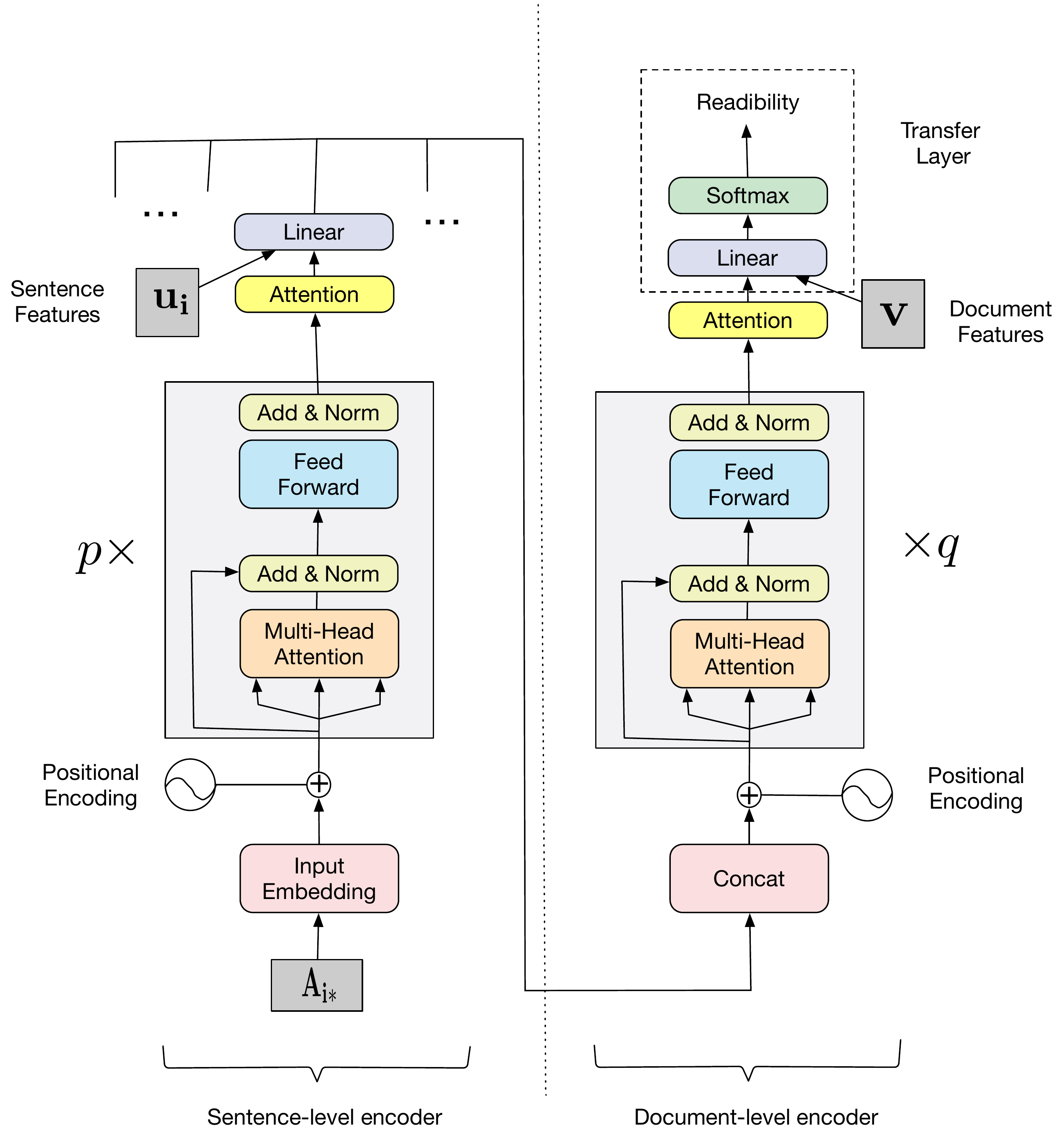}
	\caption{ReadNet: proposed hierarchical transformer model specialized for readability analysis \label{fig:model}}
	\vspace{-0.5cm}
\end{figure}

 In order to address the limitations of traditional approaches, we propose ReadNet: the Hierarchical Transformer model for readability analysis as shown in Figure~\ref{fig:model}.

The proposed model incorporates the explicit features with a hierarchical document encoder that encodes the sequence and structural information of an article. The first level of the hierarchical learning architecture models the formation of sentences from words. The second level models the formation of the article from sentences. The self-attention encoder (to be described in subsection~\ref{subsec:transformer}) is adapted from the vanilla Transformer encoder~\cite{vaswani2017attention}. The hierarchical structure, attention aggregation layer, combination with explicit features and transfer layer are specially designed for this readability analysis task.

\vspace{-0.5cm}
\subsection{From Words to Sentences}

In this subsection, we introduce the encoding process of sentences in hierarchical mutli-head self-attention. 
The encoding process has three steps: \textit{1)} the self-attention encoder transforms the input sequence into a series of latent vectors; \textit{2)} the attention layer aggregates the encoded sequential information based on the induced significance of input units; \textit{3)} The encoded information is combined with the explicit features.
\vspace{-0.6cm}
\subsubsection{Transformer Self-Attention Encoder}\label{subsec:transformer}
This encoder is adapted from the vanilla Transformer encoder~\cite{vaswani2017attention}. The input for this encoder is $\mA_{i,:}$, which represents the $i$-th sentence. 

The Embedding layer encodes each word $A_{i,j}$ into a $d$-dimensional vector based on word embedding. The output is a $m \times d$-dimensional matrix $\mB$ where $d$ is the embedding dimension and $m$ is the number of words.

The position encoding layer indicates the relative position of each word $A_{i,j}$. The elements of positional embedding matrix $\mP$ where values in the $i$-th row $j$-th column is defined as follows.	

\begin{equation}
P_{i, j}=
\begin{cases}
\sin(i / 10^{4j/d})& { j \mbox{ is even} }\\
\cos(i / 10^{4(j-1)/d})& { j \mbox{ is odd}}
\end{cases}
\end{equation}

The embedded matrix $\mB$ and positional embedding matrix $\mP$ are added 
into the initial hidden state matrix $\mH^{(0)} = \mB + \mP$.
$\mH^{(0)}$ will go through a stack of $p$ identical layers. Each layer contains two parts: (i) the Multi-Head Attention donated as function $ f_{MHA} $ defined in Equation~\ref{eq:mha}, and (ii) the Position-wise Feed-Forward $ f_{FFN}$ defined in Equation~\ref{eq:ffn}. Layer normalization is used to avoid gradient vanishing or explosion.

\emph{Multi-head Self-Attention function}($f_{MHA}$) \cite{vaswani2017attention}  encodes the relationship among query matrix $\mQ$,  key matrix $\mK$ and value matrix $\mV$ from different representation subspaces at different positions. $d_k = d/h$. $\mW$ is a $d \times d$ weight matrix. $\oplus$ denotes concatenation. $\mW_{Ki}, \mW_{Vi}, \mW_{Qi}$ are $d \times d_k$ weight matrix for head function $g_i$.  
\begin{equation}\label{eq:mha}
f_{MHA} (\mQ, \mK, \mV) = (g_1(\mQ, \mK, \mV)) \oplus  ...\oplus g_h(\mQ, \mK, \mV) )  \mW 
\end{equation}
\begin{equation}
g_i(\mQ, \mK, \mV)  = \mathrm{softmax}(\frac{ \mQ \mW_{Qi} (\mK \mW_{Ki})^{T}}{\sqrt{d_k}}) (\mV \mW_{Vi})
\end{equation}

\emph{Position-wise Feed-Forward Function} $f_{FFN}$ \cite{vaswani2017attention}  adopts two 1-Dimensional convolution layers with kernel size 1 to encode input matrix $\mX$.
\begin{equation}\label{eq:ffn}
f_{FFN}(\mX) = \mathrm{Conv1D}( \mathrm{ReLU} ( \mathrm{Conv1D}(\mX)))
\end{equation}

For the $l$-th encoder layer, $\mH^{(l)}$ is encoded into  $\mH^{(l+1)}$ according to Equation~\ref{eq:encoder}
\begin{equation}\label{eq:encoder}
\mH^{(l+1)} = f_{FFN}( f_{MHA} (\mH^{(l)} , \mH^{(l)} , \mH^{(l)}))
\end{equation}

\subsubsection{Attention Aggregation Layer}


After $p$ transformer encoder layers, each sentence $\mA_{i,:}$ is encoded into a $m \times d$-dimensional  matrix $\mH^{(p)}$.

We first pass $\mH^{(p)}$ through a feed forward layer with $d \times d$ dimensional weights $\mW_1$ and
bias term $b_1$ to obtain a hidden representation as $\mU$:
\[
\mU  = \tanh(\mH^{(p)} \mW_1 + b_1) ,
\]
then compute the similarity between $\mU$ and
the trainable $d \times 1$ dimensional context matrix $\mC$ via
\[
\vw = \mathrm{softmax} (\mU \mC)\,,
\]
which we use as importance weights to obtain the final embedding of the sentence $\mA_{i,:}$:

\begin{equation}\label{eq:att-h}
\vh_i = \sum_{byRow} \mH^{(p)} \cdot \vw
\end{equation}

\subsubsection{Combination of explicit features}
The sentence level features $\vu_i$ introduced in Section~\ref{indicator} Table~\ref{tb:features} for $i$-th sentence are concatenated by ${\vh_i^*} = {\vh_i} \oplus {\vu_i}$ .

  \subsection{From Sentences to Articles}

The second level of the hierarchical learning architecture is on top of the first layer. $n$ encoded vector ${\vh_i^*} (1 \leq i \leq n)$ are concatenated as the input for this layer. The structure of second level is the same as the first level. The output of this level is a vector $\vy$ as the overall embedding of this article.
 

\subsection{Transfer layer}

The goal of the transfer layer is to improve prediction quality on a target task where training data are scarce, while a large amount of other training data are available for a set of related tasks. 

The readability analysis problem suffers from the lack of labeled data. Traditional benchmark datasets labeled by domain experts typically contain a small number of articles. For instance, CEE contains 800 articles and Weebit contains around 8 thousand articles. 
Such quantities of articles are far smaller than those for sentiment or topic-related document classification tasks which typically involve over ten thousand articles even for binary classification \cite{CNN,chen2019subarticle}.
On the other hand, with the emerging of online encyclopedia applications such as Wikipedia, it provides a huge amount of training dataset. For instance, English Wikipedia and Simple-English Wikipedia contain more than 100 thousand articles which can be used to train a deep learning model.

One fully connected layer combines the article embedding vector $\vy$ and document-level features ${\vv}$ from Table~\ref{tb:features} to output the readability label vector $\vr$ after a Softmax function. $\mW_t$ is the weight of the fully connected layer. For dataset with $m$ categories of readability ratings, each document is embedded into $\vr$ with $m-1$ dimensions. 
\begin{align*}
\begin{split}
& {\vr} = \mathrm{softmax}({\mW_t}(\vy \oplus {\vv})) 
\end{split}
\end{align*}

If transfer learning is needed, instead of random initialization, this network is initialized with a pre-trained network based on a larger corpus. During the training process, update the transfer layer while keeping all other layers frozen. If transfer learning is not needed, all layers are updated during the training process.

\subsection{Learning Objective}
Given dataset with $m$ categories of readability ratings, the goal is to minimize ordinal regression loss~\cite{rennie2005loss} defined as Equation~\ref{eq:cross}.  $\vr_{k}$ represents the $k$-th dimension of the $\vr$ vector. $y$ is the true label. The threshold parameter $\theta_1, \theta_2, ...\theta_{m-1}$ are also learned automatically from the data.


\begin{equation}\label{eq:cross}
L(\vr;y) = -\sum_{k=1}^{m-1}f(s(k;y)(\theta_k - \vr_k)), \quad where \quad
s(k;y)=\begin{cases}
-1 &  k<y\\
+1 & k\geq y
\end{cases} 
\end{equation}

Here, the objective of learning the readability analysis model is essentially different from that of a regular document classification model, since the classes here do form a partial-order. However, the case of two classes degenerates the learning to the same as that of a binary classifer.

\subsection{Why Hierarchical Self-attention}

For self-attention, the path length in the  computation graph between long-range dependencies in the network is $O(1)$ instead of $O(n)$ for recurrent models such as LSTM. Shorter path length in the computation graph makes it easier to learn the interactions between any elements in the sequence. For readability analysis, modeling the overall interaction between words is more important than modeling the consequent words. For semantic understanding, the consequence of two words such as `` very good '' and ``not good" make distinct semantic meanings.  While for readability analysis, it does not make difference in difficulty to understand it. The overall evaluation of the words difficulties in the sentences matters. 

The hierarchical learning structure benefits in two ways. First, it mimics human reading behaviors, since the sentence is a reasonable unit for people to read, process and understand. People rarely check the interactions between arbitrary words across different sentences in order to understand the article. Second, the hierarchical structure can reduce parameter complexity. For a document with $n$ sentences, $m$ words per sentence, $d$ dimension per word, the parameter complexity of the model is $O((nm)^2 d)$ for single level structure. While for the hierarchical structure, the parameter complexity is $O(m^2d + n^2d)$.

%% file: experiment.tex
\section{Experiments}

In this section, we present the experimental evaluation of the proposed approach.
We first introduce the datasets used for the experiments, followed by the comparison of the proposed approach and baselines based on held-out evaluation, as well as detailed ablation analysis of different techniques enabled by our approach.

\subsection{Datasets}
We use the following three datasets in our experiment. Table~\ref{tb:stat} reports the statistics of the three datasets including the average number of sentences per article $n_{sent}$ and the average number of words per sentence $n_{word}$.

 \textbf{Wiki} dataset ~\cite{wikidata} contains \emph{English Wikipedia} and \emph{Simple English Wikipedia}. Simple English Wikipedia thereof is a simplified version of English Wikipedia which only uses simple English words and grammars. This dataset contains 59,775 English Wikipedia articles and 59,775 corresponding Simple English Wikipedia articles.

 \textbf{Cambridge English Exam (CEE)}~\cite{CambridgeData} categorizes articles based on the criteria of five Cambridge English Exam level  (KET, PET, FCE, CAE, CPE). The five ratings are sequentially from the easiest KET to the hardest CPE. In total, it contains 110 KET articles, 107 PET articles, 153 FCE articles,  263 CAE articles and 155 CPE articles. 
 Even though this dataset designed for non-native speakers may differ from materials for native English speakers, the difficulty between five levels is still comparable.
 We test our model on this dataset in order to check whether our model can effectively evaluate the difficulty of English articles according to an existing standard.

\textbf{Weebit}~\cite{weebit} is one of the largest dataset for readability analysis. It contains 7,676 articles targeted at different age group readers from Weekly Reader magazine and BBC-Bitesize website. Weekly Reader magazine categorizes articles according to the ages of targeted readers in 7-8, 8-9 and 9-10 years old. BBC-Bitesize has two levels for age 11-14 and 15-16. The targeted age is used to evaluate readability levels.

\begin{table*}
	\centering
	\small
	\begin{tabular}{c|cc|ccccc|ccccc}
		\multirow{2}{*}{Datasets}&\multicolumn{2}{c|}{{Wiki}}&\multicolumn{5}{c|}{Cambridge English Exam}&\multicolumn{5}{c}{WeeBit}\\
		\cline{2-13}
		&En & \scriptsize{Simple} En & KET & PET  & FCE  & CAE & CPE& WR 2 & WR 3 & WR 4 & KS3 &GCSE   \\
		\hline
		$n_{sent}$ & 37.46 &7.74 & 6.30  &8.80 & 16.47 & 10.63 & 16.69  &23.41 & 23.28 & 28.12 & 22.71 & 27.85 \\
		\hline
		$n_{word}$ & 17.03& 14.41 &9.40 & 16.63 & 17.96 & 16.39 & 23.47   &12.56 & 13.48 & 16.29 & 20.04 & 18.62\\
		\hline
	\end{tabular}
	\caption{Statistics of datasets {Wiki}, {Cambridge English Exam} and {Weebit} }\label{tb:stat}
	\vspace{-0.5cm}
\end{table*}

\input{tblAccuracy}

\begin{table}[!h]
	\centering
	\small
	\setlength\tabcolsep{1pt}
	\begin{tabular}{c|c|c|c|c|c}
		
		& KET & PET & FCE & CAE & CPE \\
		\hline
		Scores & 0.381 $\pm$ 0.078 & 0.544 $\pm$ 0.092& 0.620 $\pm$ 0.054 & 0.671 $\pm$ 0.085 & 0.837 $\pm$ 0.071 \\
		\hline
	\end{tabular}
	\caption{Average readability scores of 10 randomly selected articles in Cambridge English Test predicted by our model trained using Wikipedia.  PET, KET , FCE, CPE and CAE have increasing difficulty levels according to Cambridge English. The scores are the confidence scores of classified as regular English Wikipedia instead of simple English Wikipedia.}\label{tb:CEPP}
	\vspace{-1.0cm}
\end{table}

\subsection{Evaluation}
In this subsection, we provide a detailed evaluation of the proposed approach.

\textbf{Baseline approaches}. We compare our proposed approach (denoted ReadNet) against the following baseline methods.
\begin{itemize}
	\item Statistical classification algorithms based on explicit features: this category of baselines including the statistical classification algorithms that are widely adopted in a line of previous works \cite{ReadingMeasures,SVM,collins2005predicting,pitler2008revisiting,pilan2016predicting,CambridgeData}, such as  multi-class Logistic Regression, the Linear SVM, and the Multilayer Perceptron (MLP)~\cite{weebit}. 
Explicit features on which these models are trained have been introduced in Section~\ref{indicator}.
Since this work targets at proposing  a  more  advanced  model to utilize features instead of proposing new features, all these features from Table~\ref{tb:features} are used.

  \item Neural document classifiers: this category of baselines represents the other line of previous works that adopt variants of neural document models for sentence or document classification.
      Corresponding approaches including the Convolutional Neural Networks (CNN) \cite{CNN}, the Hierarchical Gated Neural Network with Long Short-term Memory (LSTM) \cite{tang2015document}, and the Hierarchical Attention Network (HATT) \cite{RNN}.
  	\item The Hierarchical Attention Network combined with explicit features (HATT+), for which we use the same mechanism as our proposed approach to incorporate the explicit features into the representation of each sentence by the attentive RNN.
\end{itemize}

\textbf{Model configurations.}  
For article encoding, we limit the number of sentences of each article to up to 50, zero-pad short ones and truncate over-length ones. According to the data statistics in Table~\ref{tb:stat}, 50 sentences are enough to capture the majority of information of articles in the datasets. For each sentence, we also normalize the number of words to be fed into the model as 50, also via zero-padding and truncating. We fix the batch size to 32, and use Adam~\cite{duchi2011adaptive} as the optimizer with a learning rate 0.001. The epochs of training for the neural models are limited to 300. We set the number of encoder layers $p$ and $q$ to 6. The embedding dimension $d=100$. Number of heads $h$ in $f_{MHA}$ is 3. CNN adopts the same configuration as~\cite{CNN}. Other statistical classification algorithms are trained until converge. Source code will be available in the final version.

\textbf{Evaluation protocol.} We formalize the task as a classification task following previous works on the three benchmark datasets. In order to provide a valid quantitative evaluation, we have to follow the existing evaluation method to show the advantage of our proposed model compared with the baselines. We adopt 5-fold cross-validation to evaluate the proposed model and baselines. We report the classification accuracy that is aggregated on all folds of validation. 


\textbf{Results.} The results are reported in Table~\ref{tbl:cv}.
Traditional explicit features can provide satisfying results. Since the multi-class logistic regression, SVM and MLP models can combine the features \emph{number of words per sentence} and \emph{number of syllabi per word} which are included in Flesch-Kincaid score, they provide the reasonable result.
	 CNN is only slightly better than random guess.  We assume that this is because CNN does not capture the sequential and structural information of documents. 
	 The HATT approach provides the best among models without explicit features. The reasons root in the structure of the model which is able to capture length and structural information of the article. Since it also adopted a hierarchical structure, the conciseness of each sentence and that of the overall article structure is captured, which appears to be significant to the task.
	 The explicit features further improve the results of HATT as shown by HATT+.
	 Even without explicit features, our proposed approach is better than HATT+.
     HATT has appeared to be successful at highlighting some lexemes and sentence components that are significant to the overall meanings or sentiment of a document. However, unlike topic and sentiment-related document classification tasks, readability does not rely on several consecutive lexemes, but the aggregation of all sentence components. The path length in the computation graph between arbitrary components dependencies in ReadNet is $O(1)$ instead of $O(n)$ for HATT. Shorter path length in the computation graph makes it easier to learn the interactions between any arbitrary words in sentence level, or sentences in document-level.

Compared with traditional approaches, the main advantage of the proposed approach is that it uses the document encoder to learn how words are connected into sentences and how sentences are connected into documents. Baseline approaches only use the averaged explicit features of all the sentences. For these datasets, several extremely difficult and complicated sentences usually determine the readability of a document. This useful information is averaged and weakened by the total number of sentences in baselines.

\subsection{Analysis on Transfer Learning}
As shown in Table~\ref{tbl:cv}, the standard deviation of the CEE task is large compared with those in Wiki and Weebit tasks since the quantity of CEE articles is not enough to train a complex deep learning model. Transfer layer in ReadNet is utilized in three steps. First is to train and save the model from larger datasets such as Wiki or Weebit. Then, we initialize the model for CEE task and load the parameter weights from the saved model except for the transfer layer. Eventually on the target task, the transfer layer is trained while keeping all other layers fixed. As shown in Table~\ref{tb:transfer}, loading a pre-trained model based on Weebit or Wiki can increase the accuracy and decrease standard deviation on the CEE task. It is shown that a more accurate and stable model can be achieved by utilizing the transfer layer and well-trained models from related tasks.

\begin{table}[!h]
    \vspace{-0.5cm}
	\centering
	\small
	\begin{tabular}{c|c|c|c}
		
		 & Original & Load Weebit & Load Wiki \\
		\hline
		Accuracy & 0.528 (0.045) & 0.568 (0.012)& 0.561 (0.014)\\
		\hline
	\end{tabular}
	\caption{Accuracy for CEE classification using the transfer layer. Original is the model not using transfer learning, and without loading trained weights from other dataset. \emph{Load Weebit} is to load the parameters weights trained in Weebit except the transfer layer. \emph{Load Wiki} is to load the parameters weights trained in Wiki except the transfer layer. }\label{tb:transfer}
	\vspace{-1.0cm}
\end{table}


Besides directly training and evaluating the same dataset, we also tried the model trained using Wikipedia dataset and evaluate on Cambridge English dataset. 10 articles are randomly selected from each level of Cambridge English Test. The probability of being classified as regular English Wikipedia instead of simple English Wikipedia is treated as the  difficulty score. The average difficulty scores predicted by the model are shown in Table~\ref{tb:CEPP}, which shows that our produced readability score implies correctly the difficulty of English documents for different levels of exams. A larger score indicates higher difficulty. These scores correctly indicate the difficulty levels of these exams.

%% file: tblAccuracy.tex
\begin{table*}[ht]
\small
\centering
\setlength\tabcolsep{1pt}
\begin{tabular}{c|ccc|ccc|cc}
\multirow{2}{*}{Accuracy}&\multicolumn{3}{c|}{Explicit Features}&\multicolumn{3}{c|}{Semantic Features}&\multicolumn{2}{c}{Explicit+Semantic}\\
\cline{2-9}
&Logistic&SVM&MLP&CNN&LSTM&HATT&HATT+&ReadNet\\
\hline
\multirow{2}*{Wiki}  &0.822 &0.848 &0.819&0.583 &0.849&0.877&0.898& \textbf{0.912 }\\
&\scriptsize{($\pm$0.006)}&\scriptsize{($\pm$0.008)}&\scriptsize{($\pm$0.007)}&\scriptsize{($\pm$0.035)} &\scriptsize{($\pm$0.007)}&\scriptsize{($\pm$0.007)}&\scriptsize{($\pm$0.007)}& \textbf{\scriptsize{($\pm$0.006)} }\\

\multirow{2}*{CEE}  & 0.462& 0.492&0.475& 0.277 & 0.473& 0.512& 0.513& 0.528 \\
& \scriptsize{($\pm$0.027)}& \scriptsize{($\pm$0.041)}&\scriptsize{($\pm$0.044)}& \scriptsize{($\pm$0.031)} & \scriptsize{($\pm$0.047)}& \scriptsize{($\pm$0.043)}& \scriptsize{($\pm$0.041)}& \scriptsize{($\pm$0.045)} \\

\multirow{2}*{Weebit} &0.724& 0.846&0.845& 0.635& 0.886 & 0.884 &0.902& \textbf{ 0.917 }\\
&\scriptsize{($\pm$0.007)}& \scriptsize{($\pm$0.006)}&\scriptsize{($\pm$0.006)}& \scriptsize{($\pm$0.043)}& \scriptsize{($\pm$0.005)} & \scriptsize{($\pm$0.007)} &\scriptsize{($\pm$0.006)}& \textbf{ \scriptsize{($\pm$0.006)}}\\
\hline

\end{tabular}
\caption{Cross-validation classification accuracy and standard deviation ( in parentheses ) on Wikipedia(Wiki), Cambridge English Exam (CEE) and Weebit dataset. We report accuracy on three groups of models: (1) statistical classification algorithms including multi-class logistic regression, Linear SVM and Multilayer Perceptron (MLP); (2) Three types of document classifier CNN, hierarchical GRNN using LSTM cells (LSTM), Hierarchical Attention Network (HATT); (3) Hierarchical Attention Network combined with explicit features(HATT+), and our proposed approach which combines explicit features and semantics with Hierarchical Self-Attention (ReadNet). Transfer learning is not used, and all parameters in the model are initialized randomly (Transfer learning is evaluated separately in Table~\ref{tb:transfer}).  }\label{tbl:cv}
\vspace{-0.5cm}
\end{table*}

%% file: conclusion.tex
\section{Conclusion and Future Work}
We have proposed a model to evaluate the readability of articles which can make great contributions to a variety of applications. Our proposed Hierarchical Self-Attention framework outperforms existing approaches by combining hierarchical document encoders with the explicit features proposed by linguistics. For future works, we are interested in providing the personalized recommendation of articles based on the combination of article readability and the understanding ability of the user. Currently, readability of articles only evaluate the texts of articles, other modalities such as images \cite{pezeshkpour2018embedding} and taxonomies \cite{chen2018onto} considered to improve readers' understanding. More comprehensive document encoders such as RCNN \cite{chen2019pipr} and tree LSTM \cite{tai2015improved} may also be considered.

%% file: reference.tex
\begingroup
\bibliography{ref}
\endgroup

%% file: readability.bbl
\begin{thebibliography}{10}
\providecommand{\url}[1]{\texttt{#1}}
\providecommand{\urlprefix}{URL }
\providecommand{\doi}[1]{https://doi.org/#1}

\bibitem{anderson1983lix}
Anderson, J.: Lix and rix: Variations on a little-known readability index.
  Journal of Reading  \textbf{26}(6),  490--496 (1983)

\bibitem{brown2005student}
Brown, J., Eskenazi, M.: Student, text and curriculum modeling for
  reader-specific document retrieval. In: Proceedings of the IASTED
  International Conference on Human-Computer Interaction. Phoenix, AZ (2005)

\bibitem{readability2}
Chall, J.S.: Readability: An appraisal of research and application (34) (1958)

\bibitem{chall1995readability}
Chall, J.S., Dale, E.: Readability revisited: The new Dale-Chall readability
  formula. Brookline Books (1995)

\bibitem{chen2019pipr}
Chen, M., Ju, C., Zhou, G., Chen, X., Zhang, T., Chang, K.W., Zaniolo, C.,
  Wang, W.: Multifaceted protein-protein interaction prediction based on
  siamese residual rcnn. Bioinformatics  \textbf{35}(14),  i305--i314 (07 2019)

\bibitem{chen2018neural}
Chen, M., Meng, C.P., Huang, G., Zaniolo, C.: Neural article pair modeling for
  wikipedia sub-article machine. In: ECML (2018)

\bibitem{chen2019subarticle}
Chen, M., Meng, C., Huang, G., Zaniolo, C.: Learning to differentiate between
  main-articles and sub-articles in wikipedia. In: Proceedings of the IEEE
  International Conference on Big Data (2019)

\bibitem{chen2018onto}
Chen, M., Tian, Y., Chen, X., Xue, Z., Zaniolo, C.: On2vec: Embedding-based
  relation prediction for ontology population. In: Proceedings of the 2018 SIAM
  International Conference on Data Mining. pp. 315--323. SIAM (2018)

\bibitem{coleman1975computer}
Coleman, M., Liau, T.L.: A computer readability formula designed for machine
  scoring. Journal of Applied Psychology  \textbf{60}(2), ~283 (1975)

\bibitem{CollinsThompson2004ALM}
Collins-Thompson, K., Callan, J.: A language modeling approach to predicting
  reading difficulty. In: Proceedings of the Human Language Technology
  Conference of the North American Chapter of the Association for Computational
  Linguistics: HLT-NAACL 2004 (2004)

\bibitem{CollinsSurvey}
Collins-Thompson, K.: Computational assessment of text readability: A survey of
  current and future research (2014)

\bibitem{collins2005predicting}
Collins-Thompson, K., Callan, J.: Predicting reading difficulty with
  statistical language models. Journal of the American Society for Information
  Science and Technology  \textbf{56}(13),  1448--1462 (2005)

\bibitem{coxhead2000new}
Coxhead, A.: A new academic word list. TESOL quarterly  \textbf{34}(2),
  213--238 (2000)

\bibitem{readability}
Dale, E., Chall, J.S.: The concept of readability. Elementary English
  \textbf{26}(1),  19--26 (1949)

\bibitem{de2016all}
De~Clercq, O., Hoste, V.: All mixed up? finding the optimal feature set for
  general readability prediction and its application to english and dutch.
  Computational Linguistics  \textbf{42}(3),  457--490 (2016)

\bibitem{duchi2011adaptive}
Duchi, J., Hazan, E., Singer, Y.: Adaptive subgradient methods for online
  learning and stochastic optimization. Journal of Machine Learning Research
  \textbf{12}(Jul),  2121--2159 (2011)

\bibitem{Feng2009}
Feng, L., Elhadad, N., Huenerfauth, M.: Cognitively motivated features for
  readability assessment. In: Proceedings of the 12th Conference of the
  European Chapter of the Association for Computational Linguistics. pp.
  229--237. Association for Computational Linguistics (2009)

\bibitem{franccois2009combining}
Fran{\c{c}}ois, T.L.: Combining a statistical language model with logistic
  regression to predict the lexical and syntactic difficulty of texts for ffl.
  In: Proceedings of the 12th Conference of the European Chapter of the
  Association for Computational Linguistics: Student Research Workshop. pp.
  19--27. Association for Computational Linguistics (2009)

\bibitem{fry1968readability}
Fry, E.: A readability formula that saves time. Journal of reading
  \textbf{11}(7),  513--578 (1968)

\bibitem{ReadingMeasures}
Fry, E.B.: The varied uses of readability measurement today. Journal of Reading
   (1987)

\bibitem{gibson1998linguistic}
Gibson, E.: Linguistic complexity: Locality of syntactic dependencies.
  Cognition  (1998)

\bibitem{graesser2004coh}
Graesser, A.C., McNamara, D.S., Louwerse, M.M., Cai, Z.: Coh-metrix: Analysis
  of text on cohesion and language. Behavior research methods, instruments, \&
  computers  \textbf{36}(2),  193--202 (2004)

\bibitem{gunning1969fog}
Gunning, R.: The fog index after twenty years. Journal of Business
  Communication  \textbf{6}(2),  3--13 (1969)

\bibitem{heilman2007combining}
Heilman, M., etc.: Combining lexical and grammatical features to improve
  readability measures for first and second language texts. In: Human Language
  Technologies (2007)

\bibitem{heilman2008analysis}
Heilman, M., Collins-Thompson, K., Eskenazi, M.: An analysis of statistical
  models and features for reading difficulty prediction. In: 3rd workshop on
  innovative use of NLP for building educational applications (2008)

\bibitem{wikidata}
Kauchak, D.: Improving text simplification language modeling using unsimplified
  text data. In: Proceedings of the 51st annual meeting of the association for
  computational linguistics (volume 1: Long papers). vol.~1, pp. 1537--1546
  (2013)

\bibitem{CNN}
Kim, Y.: Convolutional neural networks for sentence classification. In:
  Empirical Methods in Natural Language Processing (2014)

\bibitem{kincaid1975derivation}
Kincaid, J.P., Fishburne~Jr, R.P., Rogers, R.L., Chissom, B.S.: Derivation of
  new readability formulas for navy enlisted personnel  (1975)

\bibitem{readability3}
Klare, G.R.: The measurement of readability: useful information for
  communicators. ACM Journal of Computer Documentation (JCD)  \textbf{24}(3),
  107--121 (2000)

\bibitem{li2015hierarchical}
Li, J., Luong, M.T., Jurafsky, D.: A hierarchical neural autoencoder for
  paragraphs and documents. arXiv preprint arXiv:1506.01057  (2015)

\bibitem{li2018hierarchical}
Li, Z., Wei, Y., Zhang, Y., Yang, Q.: Hierarchical attention transfer network
  for cross-domain sentiment classification. In: Thirty-Second AAAI Conference
  on Artificial Intelligence (2018)

\bibitem{lin2015hierarchical}
Lin, R., Liu, S., Yang, M., Li, M., Zhou, M., Li, S.: Hierarchical recurrent
  neural network for document modeling. In: Proceedings of the 2015 Conference
  on Empirical Methods in Natural Language Processing. pp. 899--907 (2015)

\bibitem{louwerse2001analytic}
Louwerse, M.: An analytic and cognitive parametrization of coherence relations.
  Cognitive linguistics  \textbf{12}(3),  291--316 (2001)

\bibitem{malvern2012measures}
Malvern, D., Richards, B.: Measures of lexical richness. The encyclopedia of
  applied linguistics  (2012)

\bibitem{mc1969smog}
Mc~Laughlin, G.H.: Smog grading-a new readability formula. Journal of reading
  \textbf{12}(8),  639--646 (1969)

\bibitem{mcnamara2014automated}
McNamara, D.S., Graesser, A.C., McCarthy, P.M., Cai, Z.: Automated evaluation
  of text and discourse with Coh-Metrix. Cambridge University Press (2014)

\bibitem{mcnamara2010coh}
McNamara, D.S., Louwerse, M.M., McCarthy, P.M., Graesser, A.C.: Coh-metrix:
  Capturing linguistic features of cohesion. Discourse Processes
  \textbf{47}(4),  292--330 (2010)

\bibitem{parikh2016decomposable}
Parikh, A.P., T{\"a}ckstr{\"o}m, O., Das, D., Uszkoreit, J.: A decomposable
  attention model for natural language inference. arXiv preprint
  arXiv:1606.01933  (2016)

\bibitem{pezeshkpour2018embedding}
Pezeshkpour, P., Chen, L., Singh, S.: Embedding multimodal relational data for
  knowledge base completion. In: Proceedings of the 2018 Conference on
  Empirical Methods in Natural Language Processing. pp. 3208--3218 (2018)

\bibitem{pilan2016predicting}
Pil{\'a}n, I., Volodina, E., Zesch, T.: Predicting proficiency levels in
  learner writings by transferring a linguistic complexity model from
  expert-written coursebooks. In: Proceedings of COLING 2016, the 26th
  International Conference on Computational Linguistics: Technical Papers. pp.
  2101--2111 (2016)

\bibitem{pitler2008revisiting}
Pitler, E., Nenkova, A.: Revisiting readability: A unified framework for
  predicting text quality. In: Proceedings of the conference on empirical
  methods in natural language processing. pp. 186--195. Association for
  Computational Linguistics (2008)

\bibitem{rennie2005loss}
Rennie, J.D., Srebro, N.: Loss functions for preference levels: Regression with
  discrete ordered labels. In: Proceedings of the IJCAI multidisciplinary
  workshop on advances in preference handling. pp. 180--186. Kluwer Norwell, MA
  (2005)

\bibitem{SVM}
Schwarm, S.E., Ostendorf, M.: Reading level assessment using support vector
  machines and statistical language models. In: Proceedings of the 43rd Annual
  Meeting on Association for Computational Linguistics. pp. 523--530.
  Association for Computational Linguistics (2005)

\bibitem{senter1967automated}
Senter, R., Smith, E.A.: Automated readability index. Tech. rep., CINCINNATI
  UNIV OH (1967)

\bibitem{Si2001ASM}
Si, L., Callan, J.: A statistical model for scientific readability. In: CIKM.
  vol.~1, pp. 574--576 (2001)

\bibitem{socher2013recursive}
Socher, R., Perelygin, A., Wu, J., Chuang, J., Manning, C.D., Ng, A., Potts,
  C.: Recursive deep models for semantic compositionality over a sentiment
  treebank. In: Proceedings of the 2013 conference on empirical methods in
  natural language processing. pp. 1631--1642 (2013)

\bibitem{tai2015improved}
Tai, K.S., Socher, R., Manning, C.D.: Improved semantic representations from
  tree-structured long short-term memory networks. In: Proceedings of the 53rd
  Annual Meeting of the Association for Computational Linguistics and the 7th
  International Joint Conference on Natural Language Processing (Volume 1: Long
  Papers). pp. 1556--1566 (2015)

\bibitem{tang2015document}
Tang, D., Qin, B., Liu, T.: Document modeling with gated recurrent neural
  network for sentiment classification. In: Proceedings of the 2015 conference
  on empirical methods in natural language processing. pp. 1422--1432 (2015)

\bibitem{weebit}
Vajjala, S., Meurers, D.: On improving the accuracy of readability
  classification using insights from second language acquisition. In:
  Proceedings of the seventh workshop on building educational applications
  using NLP. pp. 163--173. Association for Computational Linguistics (2012)

\bibitem{vaswani2017attention}
Vaswani, A., Shazeer, N., Parmar, N., Uszkoreit, J., Jones, L., Gomez, A.N.,
  Kaiser, {\L}., Polosukhin, I.: Attention is all you need. In: Advances in
  neural information processing systems. pp. 5998--6008 (2017)

\bibitem{CambridgeData}
Xia, M., Kochmar, E., Briscoe, T.: Text readability assessment for second
  language learners. In: Proceedings of the 11th Workshop on Innovative Use of
  NLP for Building Educational Applications. pp. 12--22 (2016)

\bibitem{RNN}
Yang, Z., Yang, D., Dyer, C., He, X., Smola, A., Hovy, E.: Hierarchical
  attention networks for document classification. In: Proceedings of the 2016
  Conference of the North American Chapter of the Association for Computational
  Linguistics: Human Language Technologies. pp. 1480--1489 (2016)

\bibitem{readability4}
Zakaluk, B.L., Samuels, S.J.: Readability: Its Past, Present, and Future. ERIC
  (1988)

\end{thebibliography}
